# Microfluidic 3D Cell Culture: Potential Application of Collagen Hydrogels with an Optimal Dose of Bioactive Glasses

Faezeh Ghobadi [1], Maryam Saadatmand [1*], Sara Simorgh [2, 3], Peiman Brouki Milan [2, 3]

[1] Department of Chemical and Petroleum Engineering, Sharif University of Technology, Tehran, Iran

[2] Cellular and Molecular Research Center, Iran University of Medical Sciences, Tehran, Iran

[3] Department of Tissue Engineering and Regenerative Medicine, Faculty of Advanced Technologies in Medicine, Iran University of Medical Sciences, Tehran, Iran

* Correspondence: Email: m.saadatmand@sharif.edu, Tel: +98 21 6616 6466

We engineered a microfluidic platform to study the effects of bioactive glass nanoparticles (BGNs) on cell viability under static culture. We incorporated different concentrations of BGNs (1%, 2%, and 3% w/v) in collagen hydrogel (with a concentration of 3.0 mg/mL). The microfluidic chip's dimensions were optimized through fluid flow and mass transfer simulations. Collagen type I extracted from rat tail tendons was used as the main material, and BGNs synthesized by the sol-gel method were used to enhance the mechanical properties of the hydrogel. The extracted collagen was characterized using FTIR and SDS-PAGE, and BGNs were analyzed using XRD, FTIR, DLS, and FE-SEM/EDX. The structure of the collagen-BGNs hydrogels was examined using SEM, and their mechanical properties were determined using rheological analysis. The cytotoxicity of BGNs was assessed using the MTT assay, and the viability of fibroblast (L929) cells encapsulated in the collagen-BGNs hydrogel inside the microfluidic device was assessed using a live/dead assay. Based on all these test results, the L929 cells showed high cell viability *in vitro* and promising microenvironment mimicry in a microfluidic device. Collagen3-BGNs3 (Collagen 3 mg/mL + BGNs 3% (w/v)) was chosen as the most suitable sample for further research on a microfluidic platform.

**Keywords:** Microfluidic system, 3D cell culture, Collagen hydrogel, Bioactive glass nanoparticle

# 1. Introduction

Tissue engineering (TE) involves biomaterials, stem cells, and growth factors as crucial elements, and its main aim is to create an *in vitro* microenvironment that mimics the natural conditions of living tissues [1,2]. The field of TE and regenerative medicine has led to increased interest in hydrogels due to their advantageous properties, such as biodegradability and high water content, which make them suitable for cell encapsulation and mimic the native extracellular matrix (ECM) [3,4]. The widespread use of hydrogels in TE has produced encouraging outcomes in terms of promoting cell attachment, proliferation, and tissue development within a 3D microenvironment. Using natural hydrogels can enhance the effects by replicating the composition and microenvironment specific to the tissue surrounding the cells, which possess less immunogenicity and cell-conductive properties [5].

Collagen is the primary constituent of the ECM and an exceptional option for promoting cell proliferation and supporting tissue regeneration [6]. It is abundant in nature, has a specific molecular structure, a low immunological reaction index, excellent biocompatibility, hydrophilic properties, and hemostatic capability [7], making it an ideal material for TE. The main limitation of collagen hydrogels, despite their versatile form, is a lack of sufficient mechanical properties and a high degradation rate [8]. To avoid these potential disadvantages, researchers added bioactive nanoparticles to the hydrogel because they are convenient to blend with polymers for an improvement in bioactivity and mechanical strength [9]. Biomedicine recognizes bioactive glass (BG) as a versatile material with a wide range of applications, from bone tissue engineering to cancer therapy [10]. Therefore, the use of BG in collagen composites is identified as a successful approach for enhancing their mechanical properties. A noteworthy study highlights the integration of BG nanofibers into collagen, showcasing a remarkable ability to diminish infections while also stimulating tissue growth, which enhances the suitability of the resulting composite for biomedical purposes [11]. Adding BG to collagen scaffolds resulted in better elastic modulus and compression than scaffolds made of pure collagen. Additionally, the presence of BG improved the biological response [12]. The composition of the BG can be modified by adding specific ions, such as copper and strontium, to enhance the cellular response of the host. In fact, the composition of the BG also influences the bioactive response and effectiveness of collagen-BG composites [13,14]. Hong *et al*.'s (2010) research revealed the presence of hydroxycarbonate apatite after three days of immersing collagen-BG scaffolds in a simulated body fluid (SBF) solution [14]. The mineralization process in SBF impacts the mechanical characteristics of collagen-BG scaffolds, causing a shift from soft to hard tissue and thereby enhancing their ability to withstand compressive stress [15,16]. In another study, Zhou *et al*. (2017) illustrated that collagen-BG nanofibers (Si/Ca/P ratio 80:15:5) enhanced mechanical properties and promoted human dermal fibroblast cell attachment, proliferation, and the secretion of collagen type I and vascular endothelial growth factor [17]. Hydrogels consisting of collagen-BG (CaO-$P_2O_5$-

SiO$_2$) exhibited increased bioactivity in SBF, creating a more stable network by facilitating chemical interactions between BG and collagen molecules [18]. Yang *et al.* (2021) developed dual injectable composite hydrogels with copper-doped bioactive glass-ceramic microspheres (CuBGM). The 3% CuBGM group showed superior osteogenic effects and cell variability, enhancing the microenvironment for cells [19]. By researching how cells and tissues act as parts of live organs that contain several tightly opposed tissue types, it is crucial to fully appreciate the genesis, functionality, and pathology of tissues [20]. In terms of their 3D shape, mechanical characteristics, and biochemical microenvironment, these tissues are very dynamic and changing. Regrettably, most research on the regulation of cells and tissues has relied on studying cells cultivated in 2D cell culture models, which do not accurately replicate the physiological conditions of any tissues or organs or the *in vivo* cellular microenvironment [21,22]. Consequently, these cultures often lose their differentiated functions. Efforts to overcome these limitations led to the creation of 3D cell culture models that use ECM to grow cells [23]. An ideal 3D culture should promote cell growth by supplying essential nutrients, moisture, and oxygen while also removing degradation products. Conducting 3D cell culture on a plate is possible; however without fluid flow channels, its effectiveness is limited. This is due to the inability to replicate the dynamic environment required for cells, which mimics *in vivo* conditions and induces organ polarity through fluid shear stress. Despite providing a more physiological environment, the absence of human cells in animal models limits their ability to predict human tissue responses [24]. Additionally, conducting long-term studies using animal models is impractical for various reasons, such as the need for ethical clearance, species differences, and the high costs involved. The factors mentioned above result in exorbitant costs and time-consuming testing with traditional methods. As a result, scientists are investigating microfluidic models for preclinical studies as a means of addressing these shortcomings [25].

Microfluidics involves using channels ranging from tens to hundreds of micrometers in size to manipulate small fluid volumes. These microfluidic devices offer several advantages, such as a high surface-to-volume ratio, the ability to apply shear stress, and control over chemical and physical gradients, which are crucial in biological settings [26,27]. These advantages are not achievable when studying cells at the macroscale or *in vivo*. Microfluidic models have been developed to simulate tissues and organs, creating organ-on-a-chip systems that represent physiological cellular microenvironments [28]. Consequently, these microfluidic models provide a platform for creating human physiological models, drug discovery research, and toxicology studies that hold potential for replacing animal testing [29,30]. Over the past decade, there has been extensive utilization of organ-on-a-chip systems capable of replicating the physical conditions of various organs, including the lung [31], kidney [32], heart [33], intestine [34], teeth [35], and bone [36].

Some 3D biomimetic microfluidics models utilized collagen and fibrin hydrogel with hydroxyapatite (HA) for cellular studies. For example, Jeon *et al.* (2014) [37] created a microfluidic chip with two lateral media channels and an interposed gel channel, enabling 3D cell culture with functional microvascular networks to mimic pathophysiological conditions. Jusoh *et al*. (2015) [28] designed a microfluidic chip with four parallel channels separated by 100 μm gaps, utilizing fibrin hydrogel and HA nanoparticles to replicate 3D microvasculature and bone tissue environments. By adjusting the HA concentration in the ECM, they influenced its mechanical properties and improved angiogenesis, developing a model for vascularized bone tissue. Ahn *et al*. (2019) [38] created a microfluidic platform with five parallel channels to investigate interactions between the bone tumor microenvironment (TME) and HA. They developed a bone TME model for poorly understood colorectal and gastric cancer. Hao *et al*. (2018) [39] introduced a "bone-on-a-chip" model that spontaneously grows 3D mineralized bone tissue without the need for differentiation agents. This system allows the co-culture of breast cancer cells with bone tissues, providing a valuable *in vitro* platform. Additionally, they designed a microfluidic device for co-culturing osteoblast cells, breast cancer cells, and fibroblast cells in a mineralized collagen hydrogel, enabling the investigation of cancer cell progression within the collagen hydrogel.

According to our knowledge, despite the numerous properties of BGNs, their potential has largely been overlooked within the realm of microfluidics for 3D cell culture, and up to now, no one has investigated using BGNs with collagen type I hydrogel for microfluidic applications. Here, we designed and present a platform that integrates collagen-bioactive glass nanoparticles (collagen-BGNs) hydrogel into a microfluidic chip to imitate the tissue microenvironment. For this purpose, collagen type I was combined with an optimum dose of BGNs in order to mimic a real tissue matrix. Additionally, we investigated the viability of fibroblast cells encapsulated in the hydrogel, which facilitated the creation of a 3D cell culture model on a microfluidic chip **(Figure. 1)**.

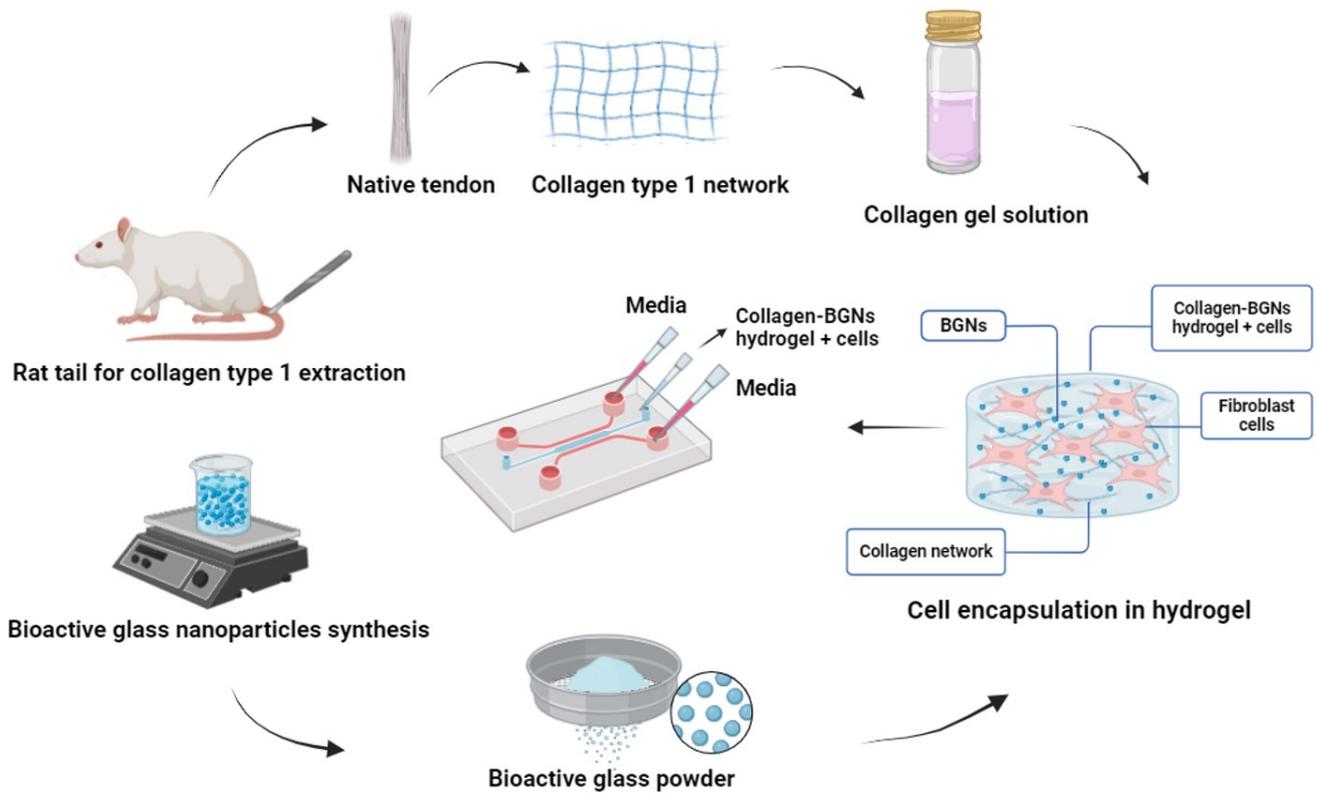

**Figure. 1**. Schematic overview of the research.

## 2. Materials and Methods

### 2.1. Microfluidic System

**(Figure. 2)** depicts the microfluidic device used in this study. It comprises two lateral channels functioning as media channels, with dimensions of 650 µm in width and 6600 µm in length. Additionally, there is a central channel measuring 900 µm in width, which serves as a host for a 3D ECM in the form of collagen type I gel. The lateral channels are interconnected with the central region of the device through the spaces between pillars. These trapezoidal posts, that dimensions and distance between them were determined by simulation, define the borders of each gel compartment, facilitating the surface tension-driven filling of hydrogel-containing cells.

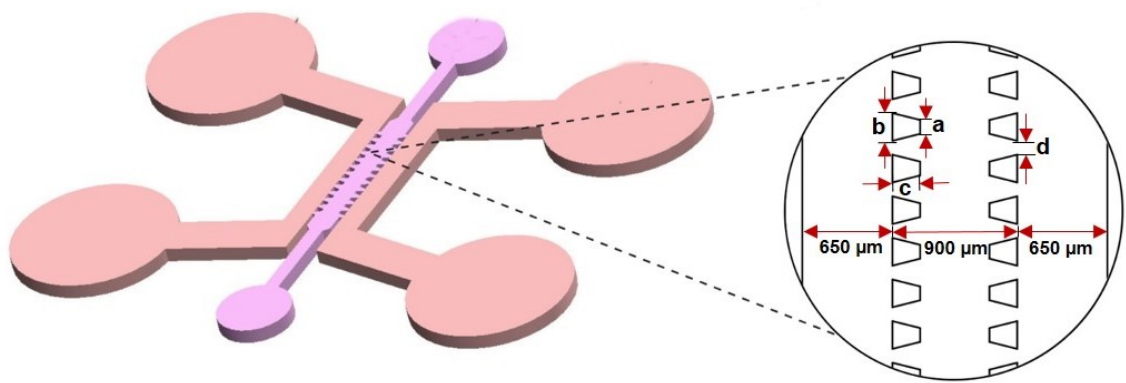

**Figure. 2**. The schematic of the microfluidic system. Its dimensions are introduced as parameters, which are explained in the simulation section. This device is composed of two lateral media channels and one interposed gel channel.

## 2.2. Numerical Simulation

Achieving a successful filling of the gels in the microfluidic gel channel relies on maintaining a balance between the capillary forces and surface tension [40]. In our system, trapezoidal posts were used between gel channels, a critical feature that allows for cell-gel and/or cell-cell interactions between adjacent compartments. These posts serve as geometric capillary burst valves, promoting optimal gel filling inside the gel channel. The creation of this interface depends on three variables: the spacing between posts, the device's surface properties, and the viscosity of the hydrogel precursor solutions. To optimize dimensions that facilitate ideal gel filling inside the microfluidic gel channel, 2D computational simulations of the gel filling process using a collagen type I solution (2.0 mg/mL) were conducted with the computational fluid dynamics (CFD) method. Initially, the geometric design of four different trapezoid posts used in the simulation was stimulated, as summarized in **(Table 1)**. The filling flow of the gel channel with collagen solution was being modeled, while the lateral channels were filled with air. Subsequently, different gel viscosities were examined due to the fact that the viscosity of the gel precursor, which is a significant factor, affects the speed at which the gel is filled (fluid motion) during the loading procedure.

The simulation involved merging the incompressible Navier-Stokes equation, the continuity equation, and the two-phase level set method, which were solved using a fixed mesh. In this investigation, the two-phase numerical analysis method employed was the level set method, as proposed by Olsson and Kreiss [41]. This approach deals with surfaces and curves on a fixed Cartesian grid, eliminating the need for parameterization of these objects. When the curve undergoes movement in the x direction, the level set equation is satisfied by its function $\varphi(x, t)$, where x represents the coordinate of a point at time t. The level set function $\varphi(x, t)$ is initialized at time $t_0$. The lateral channels are designated for the continuous phase (air, φ = 0), while the dispersed phase (collagen solution, φ = 1) fills the central channel. Therefore, the level set function can be thought of as the volume fraction of gel during the loading process. To move the interface with the velocity field u of both fluids, the modeling interface solves the following equations [42]:

Level set equation:

$$\frac{\partial \varphi_{LS}}{\partial t} + u . \nabla \varphi_{LS} = \gamma \nabla . (\varepsilon_{LS} \nabla \varphi_{LS} - \varphi_{LS}(1 - \varphi_{LS}) \frac{\nabla \varphi_{LS}}{+\nabla \varphi_{LS}} \qquad (1)$$

The left-hand side terms dictate the interface's accurate motion, whereas the right-hand side terms are critical for maintaining numerical stability. The $\varepsilon_{LS}$ parameter determines the thickness of the area where $\varphi_{LS}$ transitions smoothly from zero to one, and it's generally similar in size to the mesh elements. On the other hand, the γ parameter controls the level set function's reinitialization or stabilization amount.

Navier-Stokes equation:

$$\rho \left( \frac{\partial u}{\partial t} + u . \nabla u \right) = -\nabla p + \mu \nabla^2 u + \rho g \qquad (2)$$

Continuity equation:

$$\nabla . u = 0 \qquad (3)$$

Where $\mu$, $\rho$, $\boldsymbol{g}$, u, $p$ represent the dynamic viscosity, density, acceleration due to gravity, velocity, and pressure, respectively. The density and dynamic viscosity of the collagen solution in the channel are mentioned in **(Table 2)**. Collagen solution velocity at the inlet was defined as 50 µm/s, continued along the channel, and the filling of the gel channel was subsequently modeled. The pressure at the outlets was zero. To simulate the hydrophobic characteristics of the inner walls, a wetted wall boundary condition was implemented. The contact angle, defined as the angle between the liquid interface and the polydimethylsiloxane (PDMS) side wall, key variables that controlled the gel filling process, was set at 140×π/180 degree for this investigation [40]. Due to the hydrophobic nature of the inner surface of a PDMS microfluidic channel, the injected gel will remain contained within the gel loading channel and will not burst into the neighboring microfluidic channels. The model underwent meshing with various sizes of free triangular meshes, and an assessment of collagen solution leakage dependency on the number of mesh elements was conducted through a grid convergence study. In order to create a mesh with fewer mesh numbers, we tried the user control option and increased the number of elements from 1882 to 37270. As the number of mesh points increased, the relative error in collagen solution leakage to lateral channels decreased. Consequently, grid meshes comprising 8939 elements (element size: fine) were selected for this geometry.

**Table.1.**

The dimensions of the trapezoids used in the simulation.

| Design number | Short base of trapezoid ($\mu m$) | Long base of trapezoid ($\mu m$) | Height of trapezoid ($\mu m$) | Distance between trapezoid ($\mu m$) |
|---|---|---|---|---|
| 1 | 44 | 100 | 100 | 100 |
| 2 | 100 | 200 | 200 | 100 |
| 3 | 44 | 100 | 200 | 50 |
| 4 | 44 | 100 | 100 | 50 |

Operating under steady-state conditions is necessary for the microfluidic platform to work in a well-characterized and repeatable manner. For the computational model, we used finite element analysis and simulated the laminar flow and transport of dilute species modules to determine the time needed to reach steady-state conditions inside the microfluidic channels and to verify that a time-efficient establishment of concentration gradients was possible with the proposed device design. We performed simulations of the culture media (modeled as water) flowing along the side channels and moving through the porous collagen hydrogel (fixed in the central channel). In particular, by solving the equations at a steady state, as shown below [43], [44], we combined the mass transfer in porous media equations **(Eqs. 4, 5, and 6)** and the incompressible Navier-Stokes and continuity equations **(Eqs. 2 and 3)**.

Equations for transport of diluted species in porous media:

$$\nabla . J_i + u . \nabla c_i, \quad J_i = -D_i \nabla c_i \tag{4}$$

$$\frac{\rho}{\varepsilon_p} \frac{\partial u}{\partial t} + \frac{\rho}{\varepsilon_p} (u . \nabla) \frac{u}{\varepsilon_p} = \nabla . [-P_i + K] - \left( \mu K^{-1} + \beta \rho u + \frac{Q_m}{\varepsilon_p^2} \right) u \tag{5}$$

$$K = \frac{\mu}{\varepsilon_p} \left( \nabla u + (\nabla u)^T \right) - \frac{2\mu}{3\varepsilon_p} (\nabla . u) \tag{6}$$

Where c is the concentration species, u is the velocity, $\rho$ is the density, $\mu$ is the viscosity, D is the diffusion coefficient, $\varepsilon_p$ is the porosity, and K is the permeability coefficient in the porous media. These parameters are presented in **(Table 2)**.

For the molecular transport, we chose to simulate the diffusion of a 10 kDa dextran molecule within a 3D collagen type I gel with a concentration of 2.0 mg/mL to effectively simulate a gradient across the 3D collagen gel from lateral channels to the gel channel. The media velocity at the inlet of side channels was set to zero flow (static modeling), and the outlet pressure was set to zero. All boundaries were set as non-slip, simulating the gel portion as porous media.

**Table.2.**

Governing equations' parameters.

| Parameter | Value | Ref |
|---|---|---|
| Density of media (water) | 1000 kg/m$^3$ | 45 |
| Dynamic viscosity of media (water) | 1 mPa×s | 45 |
| Density of collagen solution (2.0 mg/mL) | 1000 kg/m$^3$ | 46 |
| Dynamic viscosity of collagen solution (2.0 mg/mL) | 6 mPa×s | 47 |
| Collagen solution velocity at the inlet | 50 μm/s | 47 |
| Diffusion coefficient for a 10 kDa dextran molecule within media at 37 °C | $9.25 \times 10^{-11}$ m$^2$/s | 48 |
| Diffusion coefficient for a 10 kDa dextran molecule within the 3D collagen type I gel (2.0 mg/mL) | $8.7 \times 10^{-11}$ m$^2$/s | 48 |
| Permeability coefficient in the collagen type I gel (porous media) | $0.25 \times 10^{-15}$ m$^2$ | 49 |
| Porosity of collagen type I gel | 0.40 | 49 |
| Contact angle between the liquid interface and the side wall | 140×π/180 | 40 |
| Surface tension of the gel channel | 0.07 N/m | 40 |

The accuracy of the numerical simulation was assessed by means of a grid convergence study. A physics-controlled meshing method was used to mesh the geometry, increasing the element size from extremely coarse (3698) to extremely finer (238214). However, we continued the simulation with the coarsest mesh, given that there were no significant changes in the results by increasing the mesh number.

## 2.3. Microfluidic Chip Fabrication

In this study, a three-channel microfluidic system comprising two lateral media channels and a central gel channel was employed [38]. We successfully fabricated the optimal chip design as determined by the fluid flow and mass transfer simulations. This device involved the use of trapezoidal posts with an inter-post spacing of 100 μm that were utilized to divide the compartments and support the hydrogel via surface tension in order to separate the 900 μm wide gel channel from the lateral media channels and promote ideal gel filling. SU-8 2050 was patterned onto a silicon wafer through the photolithography technique, and then a microfluidic system was fabricated out of PDMS using the soft lithography technique. Briefly, PDMS with an elastomer-to-curing agent ratio of 10:1 (w/w) was mixed and placed in a vacuum chamber to remove any air bubbles. Subsequently, PDMS was poured on the silicon mold and cured in the oven at 65 °C for 5 hours. Inlet and outlet ports were punched out by using the biopsy punch to create hydrogel injection ports (1 mm) and media reservoirs (4 mm). After that, the PDMS and the glass slide were treated with oxygen plasma under vacuum for 30 seconds to bond together and create a 200 μm-deep microfluidic chip. As shown in **(Figure. 3)**, by injecting dyed collagen solution, the fabricated device was examined for gel channel leakage.

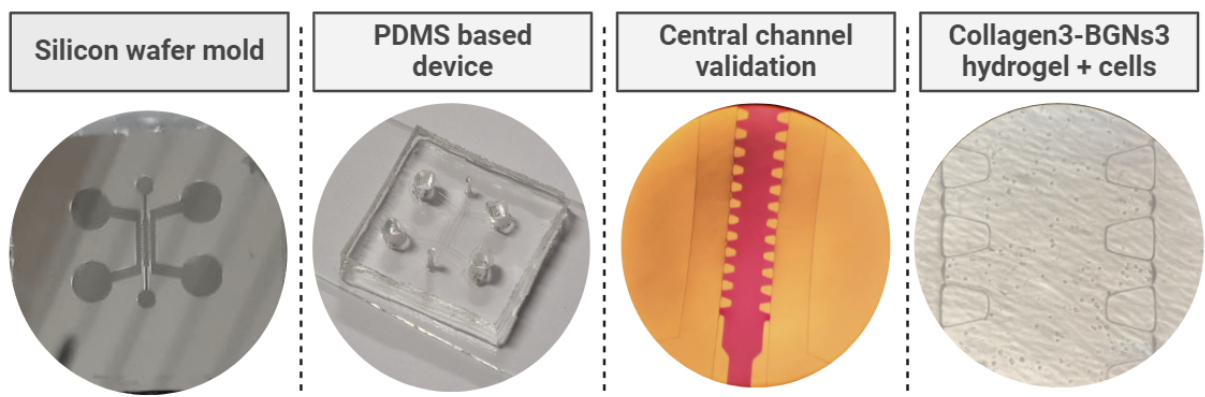

**Figure. 3.** Schematic of microfluidic chip fabrication using photolithography, collagen solution loading process without any leakage, and cells encapsulated in the collagen3-BGNs3 hydrogel into the chip.

## 2.4. Materials

Polydimethylsiloxane (PDMS, Silgard 184) was purchased from Dow Corning, MI, USA; negative photoresist (SU-8 2050) and its developer were purchased from MicroChem, USA. The following materials were purchased from Merck, Germany: ethanol, acetone, acetic acid, tetraethyl orthosilicate 99% (TEOS), nitric acid 69% ($HNO_3$), triethylphosphate (TEP), calcium nitrate 99.5%, strontium nitrate 98%, and copper nitrate 98%. The MTT cell proliferation kit was purchased from Kiazist, Iran, and the live/dead viability kit was bought from ThermoFisher, Berlin, Germany. Dulbecco's modified Eagle's medium (DMEM), fetal bovine serum (FBS), penicillin-streptomycin (pen/strep), and 0.25% trypsin-EDTA were all provided by Gibco, Canada.

## 2.5. Collagen Type I Extraction and Characterization

The rat tail tendons (from the Iran University of Medical Sciences, Tehran, Iran) were used to extract type I collagen [50]. First of all, the tendons were separated from the rat tail and collected in the petri dish containing phosphate buffered saline (PBS); then the tendons were transferred to a petri dish containing 99% acetone for 5 minutes and then transferred to 70% isopropanol for 5 minutes. Following that, the tendons were inserted in 0.02 N acetic acid and stirred with a magnetic stirrer at 4 °C for 48 h. After homogenizing the solution, it was centrifuged at 4 °C at 15,000 rpm for 45 minutes. After centrifugation, the collagen solution was filled inside the dialysis bag (MWCO: 8000), immersed in DI water, and kept stirring for 4 days at 4 °C. It was then stored at -80 °C and lyophilized to obtain a sponge. The lyophilized collagen was then dispersed in acetic acid at 0.02 N to obtain a collagen solution at the target concentrations. Hydrogel preparation will be explained in the following sections. The extracted collagen was characterized by FTIR and gel electrophoresis tests.

**Fourier Transform Infrared Spectroscopy (FTIR)**

Fourier transform infrared spectroscopy (FTIR) was used to detect amide bands in the extracted collagen. The spectra were collected at a resolution of 4 cm in 1 over 64 scans in a Nicolet 6700 spectrometer (Mundelein, IL, USA).

**Gel Electrophoresis**

Through the use of sodium dodecyl sulfate-polyacrylamide gel electrophoresis (SDS-PAGE) analysis, the purity of the extracted type I collagen was assessed. Briefly, the collagen gel sample was heated at 95 °C for 5 minutes while being combined with 125 mM Tris, 4% SDS, 10% 2-Mercaptoethanol, 20% glycerol, and 0.004% bromophenol blue. The gel was stained with 0.5% Coomassie Blue (w/v) (40% methanol, 10% acetic acid) after electrophoresis (Mini-PROTEAN Tetra Cell, Bio-Rad) [51].

## 2.6. BGNs Synthesis and Characterization

The sol-gel method was used to synthesize 58S BGNs with a composition of 58% $SiO_2$, 26% CaO, 9% $P_2O_5$, 5% SrO, and 2% CuO [52]. Briefly, 15 mL of pure ethanol and 11.125 mL of TEOS were poured into 10 mL of distilled water and stirred for 30 minutes to completely hydrolyze the mixture. At the same time, the pH of the solution was maintained between 2 and 4 via 2N $HNO_3$. Subsequently, 2.65 mL of TEP was added to the mixture, and the solution was stirred again for 30 minutes. Then 5.31 g of Ca$(NO_3)_2 \cdot 4H_2O$ was poured into the solution and stirred for 30 minutes. In the next step, 0.91 g of Sr$(NO_3)_2 \cdot 2H_2O$ and 0.39 g of Cu$(NO_3)_2 \cdot 3H_2O$ were added to the mixture; 1 M ammonia was added dropwise to the solution to increase the pH to 7.4, and the gel was formed. The obtained gel was then heat-treated at 70 °C for 48 hours, calcinated at 600 °C for 2 hours with a 10 °C/minutes heating rate, and cooled down to room temperature. Finally, to obtain a fine powder of BGNs, it was sieved through a 300-mesh sieve for further use. The BGNs synthesized were characterized by XRD, FTIR, FE-SEM/EDX, and DLS tests.

**X-ray Diffraction (XRD)**

The phase composition analysis was assessed by X-ray diffraction (XRD; PW1730, Philips, Netherlands) at a tube current of 40 mA and a generator voltage of 40 kV. The scan speed was 2°/min, and a 2Ө range of 10 to 80° was set for the diffractometer. Cu-Kα radiation was used to determine the crystal structure of the BGNs.

**Fourier Transform Infrared Spectroscopy (FTIR)**

FTIR, with a wavenumber range of 400 to 4000 $cm^{-1}$, was used to analyze the chemical structure of the BGNs. The spectra were collected at a resolution of 4 cm in 1 over 64 scans in a Nicolet 6700 spectrometer (Mundelein, IL, USA).

**Field Emission Scanning Electron Microscopy (FE-SEM)**

Field emission scanning electron microscopy with an energy-dispersive X-ray spectrometer (FE-SEM-EDX, MIRAIII, TESCAN, Czech Republic) was used to analyze the morphology and elemental composition of BGNs. Before testing, the samples were placed on an aluminum stub and sputtered with gold for 110 s at a 0.1 mbar vacuum.

**Dynamic Light Scattering (DLS)**

The particle size and the size distribution of the BGNs were investigated using dynamic light scattering (DLS, SZ100, Horiba, Japan). The BGNs were suspended in deionized water for 10 minutes before being measured at a wavelength of 632.8 nm at a temperature of 25 °C with a detection angle of 90.

**2.7. Collagen-BGNs Hydrogel Preparation and Characterization**

The collagen hydrogel was prepared at a concentration of 3 mg/mL by dissolving 9 mg of lyophilized collagen in 1300 µL of 0.5 M acetic acid under a magnetic stirrer at 4 °C until a clear solution was obtained. It was then neutralized by adding 560 µL of 1M NaOH drop by drop, and its pH was adjusted to approximately 7.4. The final neutral solution was brought to a volume of 3 mL with PBS (10X). Throughout the neutralization process, all steps were kept on ice to prevent premature gelation. Subsequently, different amounts of BGNs (0, 1, 2, and 3% w/v) were dispersed into the prepared collagen solution. The solution was then incubated at 37 °C for collagen fibrillation and hydrogel formation [53]. Experimental groups are outlined in **(Table 3)**, and the given abbreviations will be used to refer to each group in the following of this paper. The prepared hydrogel was characterized by FE-SEM and rheological analysis.

**Table.3.**

Description of the hydrogel composite samples.

| Sample Name | Collagen Concentration (mg/mL) | BGNs Concentration (w/v) % |
|---|---|---|
| Collagen3-BGNs0 | 3 | 0 |
| Collagen3-BGNs1 | 3 | 1 |
| Collagen3-BGNs2 | 3 | 2 |
| Collagen3-BGNs3 | 3 | 3 |

**Scanning Electron Microscopy (SEM)**

The morphology and microstructure of the fabricated scaffolds were assessed using scanning electron microscopy (SEM, MIRAIII, TESCAN, Czech Republic), and the cross section of freeze-dried samples was mounted on aluminum pieces and coated with gold. Using ImageJ software and SEM images of several scaffold cross-sections, the scaffold's pore size was measured. The scaffold pore size was determined by taking the mean diameter of 40–50 randomly selected pores from each image [54].

**Rheological Analysis**

The rheological properties of collagen hydrogels at different concentrations of BGNs (0%, 1%, 2%, and 3%) were examined using a rotational rheometer (Anton Paar, Graz, Austria) with a parallel plate (25 mm diameter) and a gap height of 1 mm. A strain sweeping mode was employed to measure the storage modulus (G') and loss modulus (G") at 37 °C at a fixed frequency of 1 rad/s.

**2.8. 3D Cell Culture on a Plate and MTT Assay**

In order to assess the biocompatibility and cytotoxicity of collagen-BGNs hydrogel, the MTT assay was carried out. For this purpose, L929 cell lines (obtained from the Pasteur Institute of Iran) were encapsulated in 100 μL sterilized hydrogel/well of a 96-well plate at a density of $5 \times 10^3$ cells/well (each well in the 96-well plates has a volume of 200 μL, a diameter of 6.94 mm, and a height of 5.27 mm) [55]. After 1, 3, and 7 days, a MTT kit was used. The media was removed, and a mixture of 10 μL of MTT reagent and 100 μL of serum-free media was added to each well. The media was removed after a 4-hour incubation period in a dark environment at 37 °C with 5% $CO_2$ and 95% relative humidity. Next, 100 µL of solubilized solution was added to each well, and the plate was wrapped in foil and shaken on an orbital shaker for 15-20 minutes to fully dissolve the MTT formazan. Subsequently, the measurement of absorbance at a wavelength of 570 nm was performed using a microplate reader (Bio-Rad Laboratories in Hercules, CA). The 2D cell culture media was used as a control sample. Cell viability was estimated using **(Eq. 6)** [56].

$$viability\% = \frac{\text{Mean OD}_{sample}}{\text{Mean OD}_{control}} \times 100 \tag{6}$$

## 2.9. Cell culture, loading on the Chip and Viability Analysis

L929 cells were mixed in DMEM supplemented with 10% FBS, 1% pen-streptomycin, and 1% L-glutamine, then cultured in a cell incubator at 37 °C with 5% $CO_2$. Prior to cell loading, the microfluidic chip underwent rinsing with 90% ethanol, autoclaving, and sterilization via ultraviolet radiation for 30 minutes. Subsequently, the chip was placed in a sterile, dry environment to facilitate residual water evaporation. The L929 cells were encapsulated in the collagen3-BGNs3 hydrogel solution at a cell density of $3 \times 10^3$ cells/mL. First, the collagen3-BGNs3 gel solution with cells was manually and slowly loaded into the gel channel with a 10 µL micropipette until the solution just filled the channel. After filling the gel channel, the chip was kept in an incubator set at 37 °C and 5% $CO_2$ for 20 minutes to allow the collagen to gel. Then, 20 µL of culture media was perfused through the each lateral microchannels. Ultimately, the chip was placed in a cell incubator for static culturing, and the culture media was replaced once a day. The live/dead assay of L929 cells encapsulated in the collagen3-BGNs3 hydrogel in the chip was performed using fluorescein diacetate (FDA) and propidium iodide (PI) [57]. The media was briefly withdrawn from the channels, and the culture chamber on the microfluidic chip was cleaned three times with PBS (1X) solution, followed by applying a 1:1000 dilution of FDA solution in PBS (1X) to the cells and incubating them at 37 °C for 15 min. Subsequently, the FDA solution was removed, excess stain was washed off with PBS (1X), and the samples were treated with 20 g/mL of PI at room temperature for 2 minutes. Observation of cell encapsulation in the hydrogel was performed using the Olympus IX70 fluorescence microscope. In this assay, FDA-stained living cells were fluorescent green, while PI-dyed dead cells were fluorescent red within the hydrogel.

## 2.10. Statistical Analysis

The test outcomes were reported as mean ± standard deviation (Mean ± SD). The statistical analysis was conducted with the aid of Graphpad Prism version 9 software and Two-Way Anova with Tukey's post-test.

Each test was performed at least three times. A significance level of P < 0.05 was used, where *(P < 0.05), ** (P < 0.01), and *** (P < 0.001) indicate levels of significance.

## 3. Results and Discussions
### 3.1. The Optimal Dimensions of the Chip by Simulation

The chip design incorporates a single gel channel, two side channels, and an array of post structures that effectively segregate them. The primary purpose of these post structures is to prevent gel solution leakage into adjacent channels, ensuring distinct gel compartments. Post structures provide a means of controlling the interface between gel compartments and main channels, enabling precise control over the diffusion and movement of molecules and cells. Unlike solid barriers, post structures allow for variable interface areas as required. This approach enables the association of independent gels with controlled dimensions. Successful gel filling relies on a delicate balance of capillary forces and surface tension within the microfluidic gel channel. Previous work by P. Huang *et al*. [40] revealed that hydrophilic surfaces led to gel or fluid leakage into neighboring channels upon fluid injection. Conversely, hydrophobic surfaces allowed the entire length of the gel channel to be filled without leakage. In another study, Lee *et al*. [58] illustrated the computational simulation depicting the sequential filling process of gel within the vessel channels. The hydrophobic inner channel surface, coupled with closely spaced posts, ensures a controlled filling process, preventing gel leakage into adjacent channels. Fluid flow simulation with various gap spacing (ranging from 44–100 µm) and hydrophobic surface properties were conducted to optimize this interface, ensuring the structural integrity of gels within each individual gel channel (different designs were introduced in **(Table 1)**). Gap spacing up to 100 µm was identified as effective in containing gels within a 900 µm-wide gel channel. According to the obtained results, if the distance between the posts is more than 100 µm, leakage occurs. The suitable dimensions of the trapezoidal posts for the movement of the collagen solution without leakage are design numbers 2, and 3, as depicted in **(Figure. 34a-b)**, and design numbers 1 and 4 were unsuitable for constructing the microfluidic device due to the collagen solution leakage observed in the simulation results. Therefore, for the experimental part of the study, we focused on design number 2 (a: 100 µm, b: 200 µm, c: 200 µm, and d: 100 µm), given that distance between posts is important for transferring molecules and growth factors to the cells in gel channel.

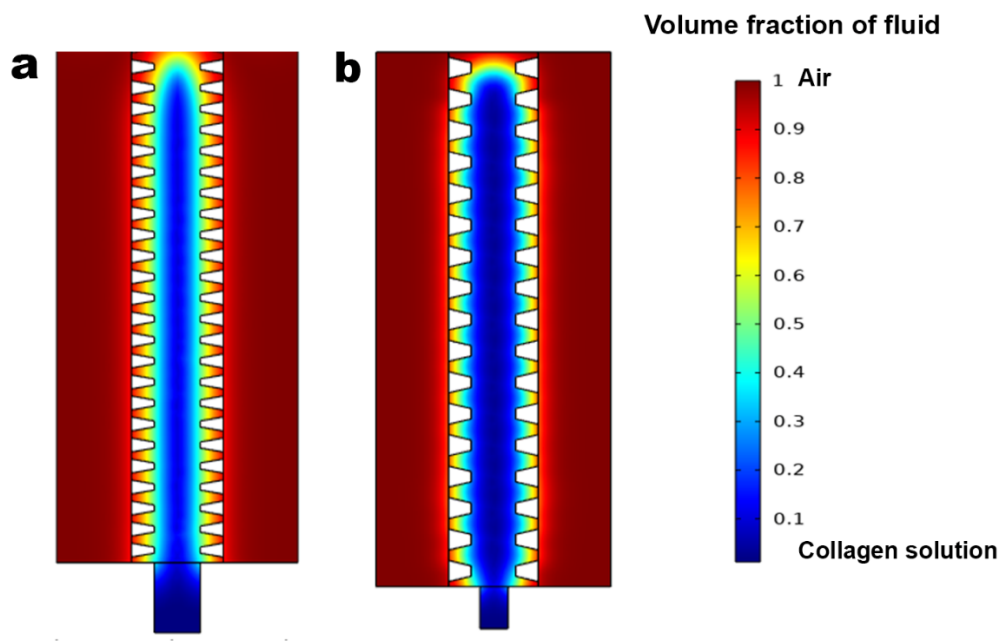

**Figure. 4.** Simulation of gel filling into the microfluidic device for two different designs (a) Movement of collagen solution without leakage along the gel channel of the design number 3; (b) Movement of collagen solution without leakage along the gel channel of the design number 2.

Ultimately, it was found that the viscosity of the hydrogel plays a crucial role in improving the connection between gel channels. Denser hydrogels exhibited higher viscosity, which increased the chance of leakage between adjacent channels. Collagen gels at concentrations of up to 6 mg/mL were effectively incorporated into the system with negligible leakage, which was less than 30%. Nevertheless, the injection of collagen gels at a concentration of 8 mg/mL was not feasible due to the potential risk of compromising the integrity of the interface. [40].

As depicted in **(Figure. 5a-b)**, the simulation results indicate that the higher the viscosity, the greater the extent of collagen solution leakage into the lateral channels. Therefore, according to these observations, collagen with a density of 3 mg/mL was used for experimental work.

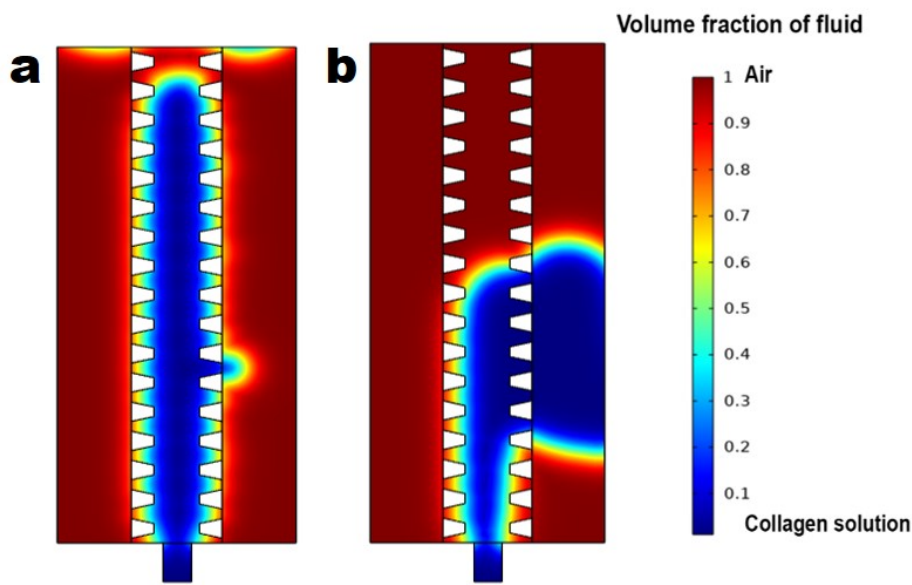

**Figure. 5.** Simulation of gel filling into the microfluidic device for different viscosities: (a) Viscosity of 9 mPa×s and low leakage of collagen solution along the gel channel; (b) Viscosity of 11 mPa×s and leakage of collagen solution and entering the media channel.

Usually, the hydrogel-containing cells are confined within a microfluidic device consisting of arrays of microposts. This setup facilitates the cultivation of cells in 3D networks while allowing for the controlled flow of conditioning fluids or culture media sideways [45]. Herein, a numerical simulation was run to confirm that the device with a 100 μm gap between posts (design number 2) can create a stable concentration within the central channel. The diffusion of a 10 kDa dextran molecule was simulated and showed that a stable concentration was established after around 4 hours within the hydrogel **(Figure. 6a-b)**. Pagano *et al*. (2014) focused on creating a stable concentration gradient within a hydrogel by investigating the diffusion of a 10 kDa dextran molecule and taking into account the device's geometric features. They observed that altering the pillar spacing in their channels led to gradient establishment times of approximately 5 hours and 4 hours, respectively, highlighting the impact of pillar spacing on gradient characteristics [45].

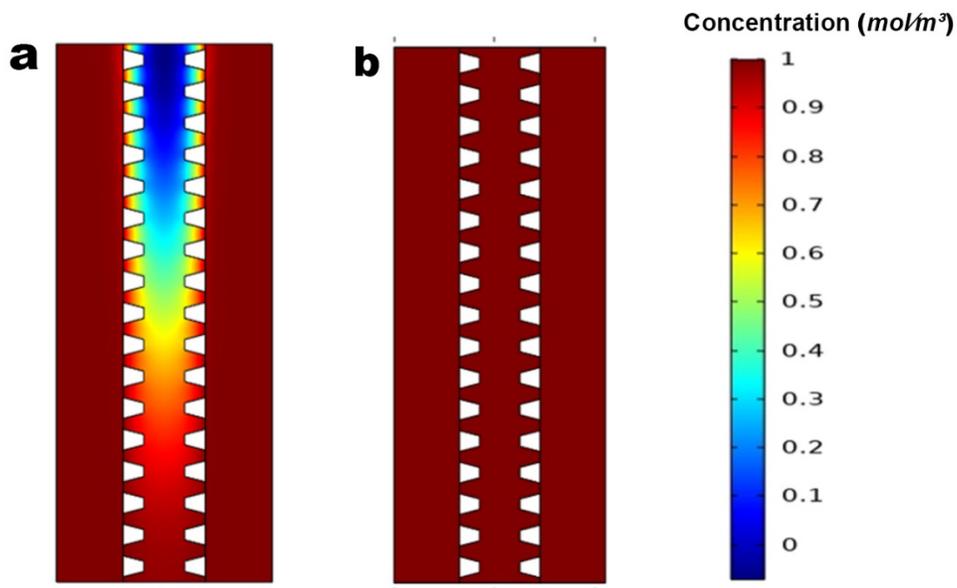

**Figure. 6.** Mass transfer simulation within the gel channel of the designed chip (Design Number 2). (a) t = 2.5 h (b) t = 4 h.

## 3.2. Characterization of Collagen Type I
**FTIR Analysis**

Collagen type I was extracted from rat tail tendons, lyophilized, and subsequently dissolved in acetic acid at a concentration of 3 mg/mL **(Figures. 7a–b)**. Under a pH of 7.4 at 37 °C, the collagen underwent fibrillogenesis (self-assembly) **(Figure. 7c)**. *In vivo*, collagen fibril formation occurs through self-assembly. However, when studied *in vitro*, this process is influenced by factors such as temperature, pH, and the presence of chemical or biological agents [59]. Developing methods to analyze fibrillation processes allows the structure and size of the resulting fibrils to be adjusted and controlled. FTIR analysis has been a longstanding method used to study the structure of collagen [60]. The FTIR spectrum of the extracted collagen is depicted in **(Figure. 7d)**, revealing discernible peaks corresponding to amides I, II, and III, as well as amides A and B. The obvious peaks at 3286 cm$^{-1}$ and 2923 cm$^{-1}$ indicate the amide A and B bands, respectively. These peaks proved that the N-H group of peptides is involved in hydrogen bonding. The collagen configuration can be determined based on the wave numbers associated with the amide I, II, and III bands. The position of the amide I bond of rat tail collagen was observed at 1625 cm$^{-1}$, indicating the presence of C=O bonds. Meanwhile, the peaks at numbers 1526, 1443, and 1394 cm$^{-1}$ are the absorption regions of amide II. The wavenumber of 1526 cm$^{-1}$ indicates the presence of N-H bonds, while the wavenumbers of 1443 and 1394 cm$^{-1}$ also indicate the absorption regions of amide II, which have $CH_2$ and $CH_3$ bonds. The amide III band has a wavenumber of 1232 cm$^{-1}$. According to the form of the amide I bond (C=O), these reflect amide II (N-H stretching and N-C deformation) and amide III (N-C deformation and N-H stretching) [61]. In essence, the FTIR spectrum vividly illustrates intact amino acids and collagen molecules. Drawing upon the work of Simorgh *et al*. (2021), these findings support the outcomes derived from FTIR analysis and thereby demonstrate the existence of certain molecular vibrations such as amide A, amide B, amide I, amide II, and amide III. These vibrational characteristics serve as compelling evidence supporting the existence of a type I collagen structure [62].

**Gel Electrophoresis**

To confirm the purity and identity of type I collagen, the SDS-PAGE analysis result is illustrated in **(Figure. 7e)**. Type I collagen is comprised of β (215 kDa), $α_1$ (130 kDa), and $α_2$ (115 kDa) chains [63]. In this study, the collagen extracted displayed two distinct bands, corresponding to distinct α-chain and β-chain types. The $α_1$ type I collagen exhibited a molecular weight of approximately 135 kDa, while the $α_2$ type I collagen measured around 110 kDa. The β chains were found to be larger than 180 kDa. Furthermore, the ratio of the $α_1$ band exceeded that of the $α_2$ band, consistent with the known composition of type I collagen's triple helix, which comprises two $α_1$ chains and one $α_2$ chain. These findings indicate that the extracted collagen aligns with the molecular weight observed in prior studies and attest to its purity [51,62].

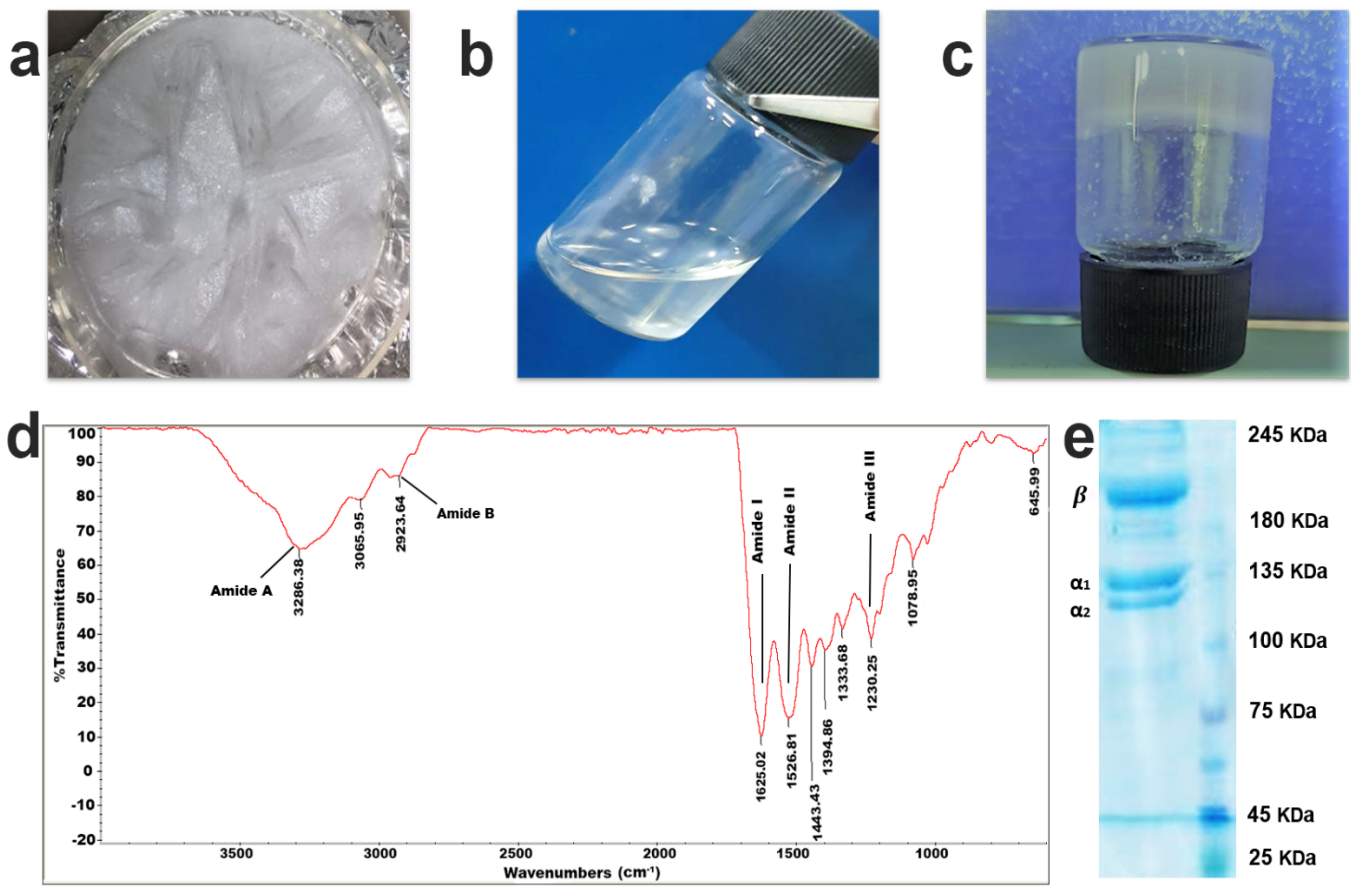

**Figure. 7.** Characterizing collagen type I, derived from rat tail. (a) Freeze-dried collagen; (b) the collagen solution, pre-crosslinking at a concentration of 3 mg/mL; (c) self-assembled collagen under pH 7.4 and 37 °C conditions; (d) The FTIR spectrum indicating specific molecular vibrations such as amide A, amide B, amide I, amide II, and amide III, confirming the presence of the type 1 collagen structure; (e) SDS-PAGE data revealed collagen type 1 with two $α_1$ chains and one $α_2$ chain, as well as a β-dimer, further confirming the structure of collagen type I.

## 3.3. Bioactive Glass Characterization

### X-ray Analysis

Examining the XRD pattern illustrated in **(Figure. 8a)**, it becomes apparent that the peaks associated with BGNs lack a crystalline structure. The XRD image serves to confirm the amorphous structure of BGNs, as it distinctly displays the broad hump characteristic of the amorphous silicate phase within the 2θ range of 25 to 35°, as reported in Ref. [64]. Upon general observation, the powders displayed a distinct blue hue, as depicted in **(Figure. 8b)**, which serves as a characteristic feature of BGs doped with copper [65].

**FTIR Analysis**

The FTIR spectrum outcomes for the sol-gel synthesized BGNs are presented in **(Figure. 8c)**. A noteworthy broad absorption peak at 1098 cm$^{-1}$ corresponds to the asymmetric stretching oscillation of the (Si-O-Si) bond. Additionally, a subtle peak around 805 cm$^{-1}$ aligns with the symmetric stretching vibration of the Si-O bond. Furthermore, a discernible peak within the 3458 cm$^{-1}$ range signifies the vibration of distinct OH groups. The presence of a low-intensity peak at 573 cm$^{-1}$ indicates the bending oscillation of the (P-O) bond, a clear indication of phosphate's incorporation as a network element in the BGNs. An absorption peak around 1646 cm$^{-1}$ is attributed to molecular water on the Si-OH (silanol) group surfaces. The FTIR spectrum of the synthesized BGNs indicates that the resulting powder possesses a BG structure [66,67].

**BGNs Morphology and Particle Size Assessment**

The morphology and particle size of BGNs were illustrated through FE-SEM analysis. As depicted in **(Figure. 8d)**, the BGNs exhibit a wide range of spherical particles with an average size of 30–60 nm, along with irregularly shaped particles. This variation indicates that the glass particles lack uniformity in morphology. However, employing DLS analysis, the average particle size of BG powders was approximated at around 100 nm **(Figure. 8e)**. This observation suggests the presence of heterogeneous BG particles due to BGNs aggregation under aqueous conditions. Also, to investigate the chemical composition of the BG powders, an EDX elemental mapping analysis was utilized. As shown in **(Figure. 8f)**, EDX mapping confirms the existence of BGNs components generated during the sol-gel process within the glass structure. The image clearly displays the presence of silicon, calcium, phosphorus, copper, and strontium elements in the glass composition.

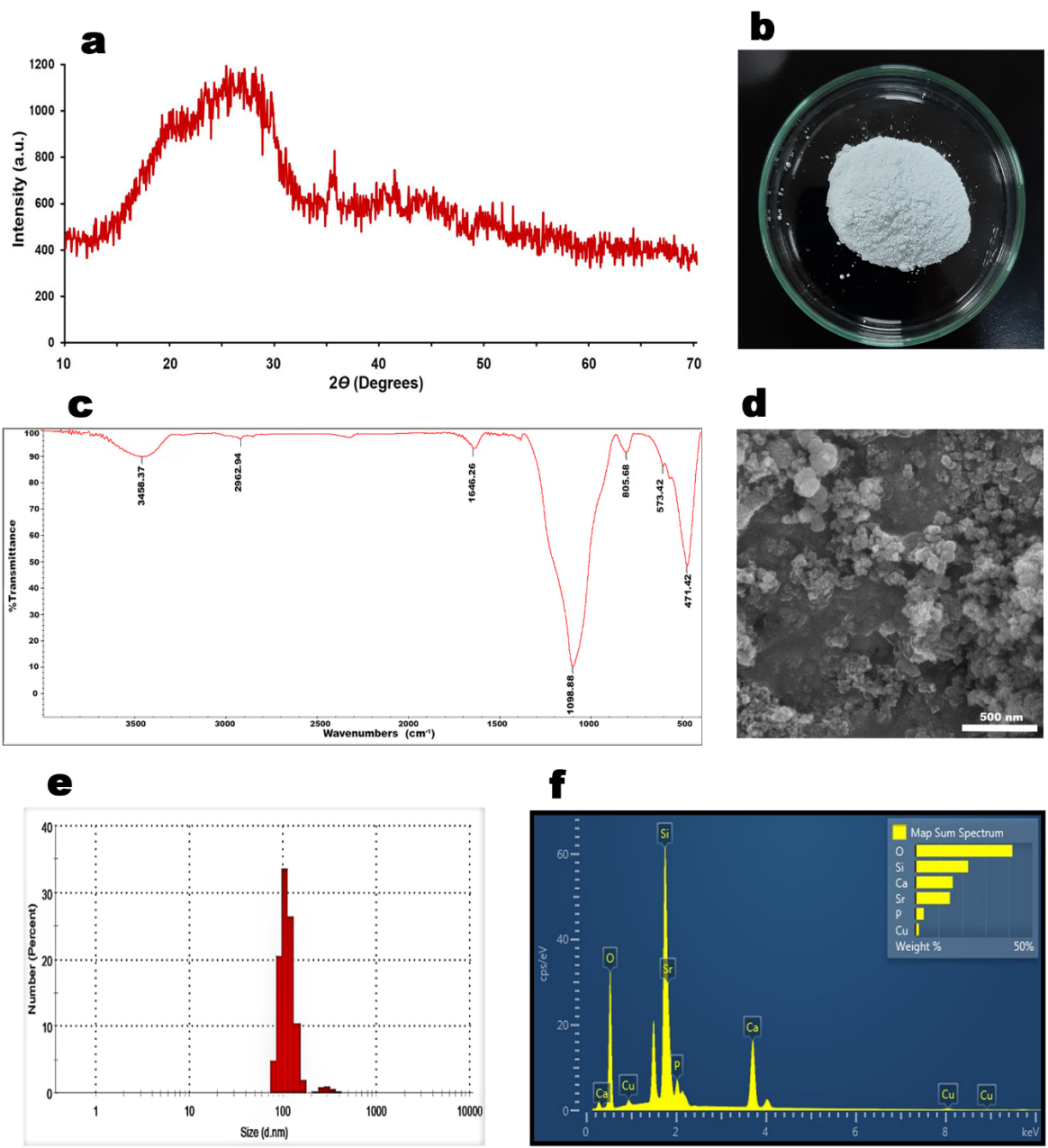

**Figure. 8.** The characterization results of the synthesized BGNs with the sol-gel method (a) XRD spectra; (b) General observation photographs; (c) FTIR spectra; (d) FESEM micrograph image of BGNs; (e) Particle size distributions.; (f) EDX analysis of BGNs.

## 3.4. Hydrogel Characterization

**Morphology of the Collagen-BGNs Hydrogel**

The structure and pore size of the collagen hydrogels at concentrations of 0, 1, 2, and 3% (w/v) BGNs following the freeze-drying procedure were observed by SEM **(Figure. 9a-f)**. The dried hydrogels possessed a highly porous structure with an average pore size (~50 μm to 250 μm), representing a typical porous structure for freeze-dried hydrogels. The incorporation of BGNs did not significantly influence the 3D structure of hydrogel [54,68].

**Mechanical Properties of the Collagen-BGNs Hydrogel**

The mechanical properties, including storage modulus (G') and loss modulus (G"), of the collagen hydrogel at varying BGNs concentrations were assessed using gel rheology under strain sweep conditions. G′ and G″ remain consistent regardless of strain within the linear viscoelastic region (LVE). The functions of G' and G" exhibit constant plateau values within this LVE range. Notably, G' surpasses G" within the LVE zone, indicating a gel-like structure of the sample. The incorporation of BGNs into collagen hydrogel notably enhances both G′ and G″ of the hydrogels. For instance, G′ of collagen3-BGNs0 starts at 64.7 Pa, escalating to 178 Pa in collagen3-BGNs1 (3 times increase), and further rising to 761 Pa in collagen3-BGNs3 (12 times increase). Similarly, G″ of collagen3-BGNs0 (17.8 Pa) experiences an elevation to 104 Pa in collagen3-BGNs3 (6 times increase). These findings suggest that the BGNs addition has a more pronounced impact on the elastic behavior of hydrogels in comparison to their viscous behavior. Nonetheless, as the strain is further increased to 100%, both G' and G" experience a significant decline, indicating liquefaction of the hydrogels. The stress sweep plots of the gels indicate a somewhat mechanically robust network for collagen3-BGNs0 (tolerance strain 21%) compared to their counterparts (26% in collagen3-BGNs1, 40% in collagen3-BGNs2, and 55% in collagen3-BGNs3). Lee *et al*.'s research validates our conclusions, demonstrating that introducing BGNs had a substantial impact on the rheological properties of the collagen hydrogel. When 200 μg/mL of BGNs was integrated into the hydrogel matrix, G' of the hydrogel experienced a nine-fold increase [69]. In summary, the incorporation of BGNs effectively enhances the mechanical properties of the hydrogels, as evidenced by the G' and G" results depicted in **(Figure. 9g)**.

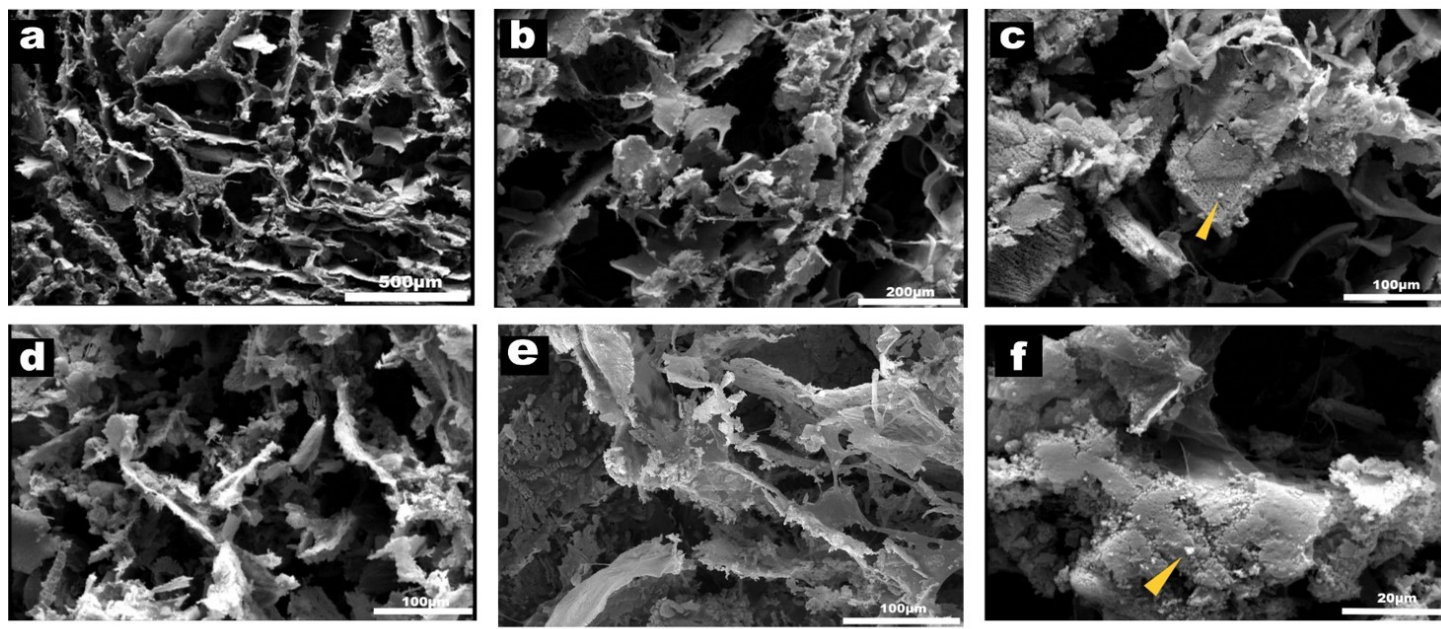

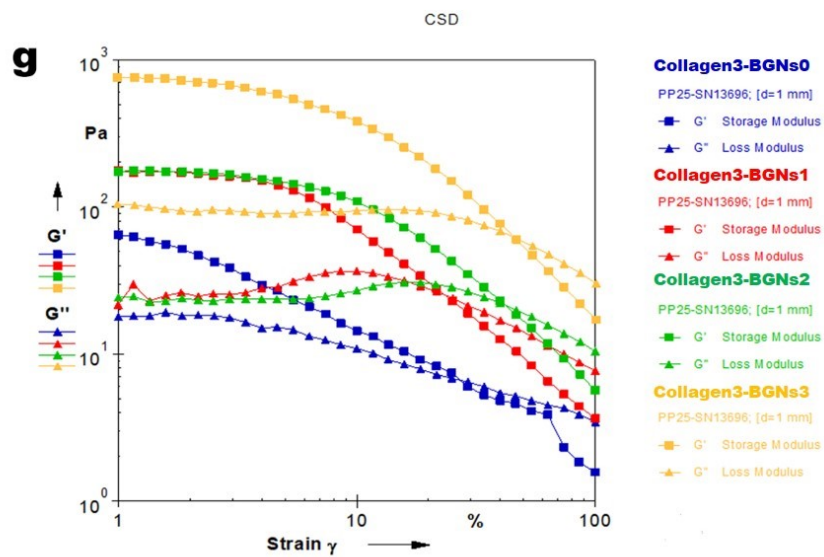

**Figure. 9.** SEM images of collagen hydrogels with varying concentrations of BGNs components. (a). collagen3-BGNs0; (b, c). Collagen3-BGNs1; (d). Collagen3-BGNs2; (e, f). Collagen3-BGNs3; Yellow arrows indicate BGNs. (g) Rheological evaluation of storage and loss modulus of different hydrogels as a function of strain at the same temperature.

## 3.5. Cell Viability

The MTT results showed that the viability of L929 cells in 2D conditions on days 1 and 3 was not significantly different according to BGNs concentration **(Figure. 10a)**. However, on day 7, cell viability in BGNs 2 and 3% was significantly increased compared to BGNs 0 and 1%. After a comprehensive assessment of different hydrogel characteristics, it is evident that collagen3-BGNs3 stands out with the highest cell viability. This phenomenon can be attributed to the ion release facilitated by BGNs, particularly those incorporating $Sr^{2+}$ and $Cu^{2+}$. A preceding study [70] has convincingly shown that these ions positively impact cell viability and proliferation, in addition to contributing to the hydrogel's favorable mechanical properties. According to Li *et al*. (2021) [71], hydrogels incorporating 2% BG exhibited favorable biocompatibility, displaying comparable living cell numbers and absorbance to the control group. In contrast, 5% BG resulted in diminished cell viability. Additionally, Moreira *et al*. (2016) [72], in their study, revealed that cells in direct contact with hydrogels containing 1% and 2% BG demonstrated no noteworthy variance in viability when compared to the control group. Thus, collagen3-BGNs3 is considered the most suitable hydrogel for utilization in the microfluidic platform. Furthermore, on the third day, the survival of cells within the microfluidic device was examined in 3D cell culture conditions. The live/dead result is shown in **(Figure. 10b)**, and the L929 cells exhibited a high cell viability rate, indicating that the materials and culture conditions employed in the model did not cause any cell death. The presence of green fluorescence indicates that most of the cells maintained their viability and exhibited ongoing proliferation within the collagen hydrogels in the chip, while red fluorescence indicates a minimal number of dead cells during the culture period. Moreover, literature suggests that cells cultured in collagen hydrogels exhibit a rounded morphology [73]. This shape is typically observed in microfluidic devices that provide a favorable environment and indicate the biocompatibility of the hydrogel.

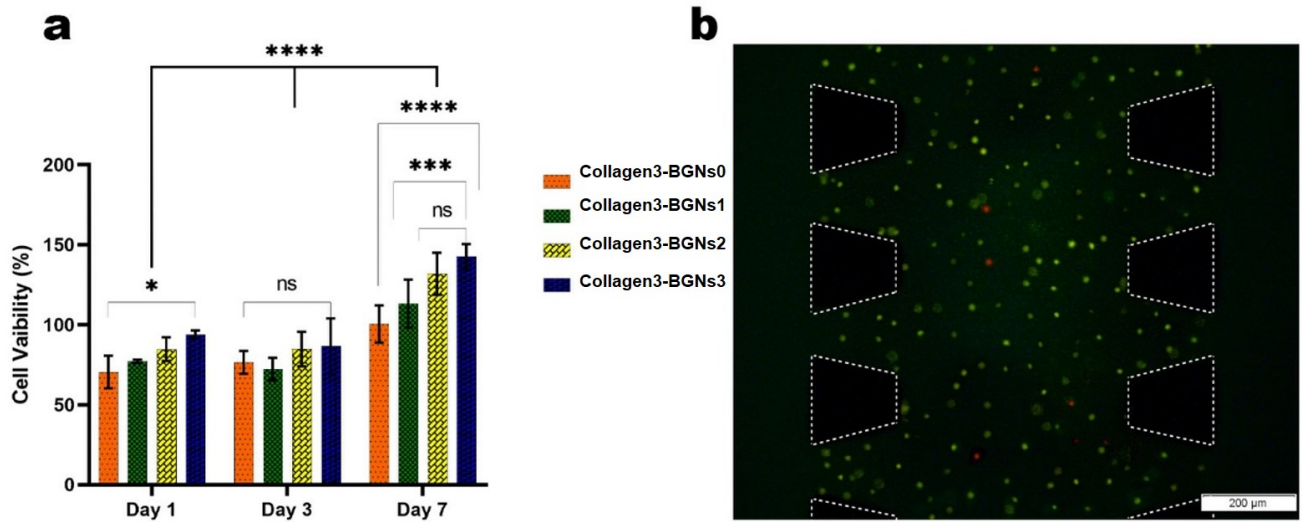

**Figure. 10.** Evaluation of cell viability encapsulated in hydrogels. (a) Comparison of the cell viability in different concentrations of BGNs by MTT assay; (NS, no significant difference; p ≤ 0.05, *** p ≤ 0.001, **** p ≤ 0.0001); (b) Live/dead was utilized to assess the viability of the cells encapsulated in collagen3-BGNs3 hydrogel into the microfluidic on day 3, the cells were highly viable, the green color shows the living cells, and the red color indicates the dead cells (scale bar = 200 μm).

## 4. Conclusion

This paper describes an approach to engineering 3D microenvironments for investigating cell viability using a microfluidic platform. The approach relies on three parallel microfluidic channels separated by an array of posts. Through fluid flow and mass transfer simulations, the chip was optimized, and trapezoidal posts with appropriate dimensions—a small base of 100 μm, a large base of 200 μm, a height of 200 μm, and a distance of 100 μm between two trapezoids—were used for fabricating the chip. Additionally, the viability of fibroblast cells was investigated both *in vitro* and within the microfluidic chip with collagen-BGNs. In summary, we utilized BGNs as a promising material to enhance the mechanical properties of the tissue-engineered scaffold. The hydrogel was primarily created using collagen type I as the ECM, and rheological testing was performed to assess its properties. The results demonstrated that the addition of BGNs had a beneficial impact on the mechanical characteristics of the hydrogel. *In vitro* studies determined the optimal concentration of BGNs based on cell viability measurements. Notably, collagen3-BGNs3 demonstrated superior cell viability compared to other BGNs concentrations. We also characterized the collagen3-BGNs3 hydrogel for L929 cell culture within the microfluidic platform, and L929 cells exhibited a high cell viability rate. In conclusion, this microfluidic chip offers a new approach for investigating differentiation applications and tissue engineering purposes, as well as bone regeneration for analyzing drug responses and toxicities in bone tissues.

## Data Availability

The datasets generated during and/or analysed during the current study are available from the corresponding author on reasonable request.

63. Amirrah, I. N. *et al.* A Comprehensive Review on Collagen Type I Development of Biomaterials for Tissue Engineering: From Biosynthesis to Bioscaffold. *Biomedicines* **10**, (2022).

64. Alasvand, N., Behnamghader, A., Milan, P. B. & Mozafari, M. Synthesis and characterization of novel copper-doped modified bioactive glasses as advanced blood-contacting biomaterials. *Mater. Today Chem.* **29**, (2023).

65. Wang, H. *et al.* Evaluation of borate bioactive glass scaffolds as a controlled delivery system for copper ions in stimulating osteogenesis and angiogenesis in bone healing. *J. Mater. Chem. B* **2**, 8547–8557 (2014).

66. Costa, H. S. *et al.* Sol-gel derived composite from bioactive glass-polyvinyl alcohol. *J. Mater. Sci.* **43**, 494–502 (2008).

67. Doostmohammadi, A., Monshi, A., Fathi, M. H. & Braissant, O. A comparative physico-chemical study of bioactive glass and bone-derived hydroxyapatite. *Ceram. Int.* **37**, 1601–1607 (2011).

68. Zare Jalise, S., Baheiraei, N. & Bagheri, F. The effects of strontium incorporation on a novel gelatin/bioactive glass bone graft: In vitro and in vivo characterization. *Ceram. Int.* **44**, 14217–14227 (2018).

69. Lee, J. H., El-Fiqi, A., Han, C. M. & Kim, H. W. Physically-strengthened collagen bioactive nanocomposite gels for bone: A feasibility study. *Tissue Eng. Regen. Med.* **12**, 90–97 (2015).

70. Alasvand, N. *et al.* Copper / cobalt doped strontium-bioactive glasses for bone tissue engineering applications. *Open Ceram.* **14**, 100358 (2023).

71. Li, J. *et al.* Bioactive nanoparticle reinforced alginate/gelatin bioink for the maintenance of stem cell stemness. *Mater. Sci. Eng. C* **126**, 112193 (2021).

72. Moreira, C. D. F., Carvalho, S. M., Mansur, H. S. & Pereira, M. M. Thermogelling chitosan-collagen-bioactive glass nanoparticle hybrids as potential injectable systems for tissue engineering. *Mater. Sci. Eng. C. Mater. Biol. Appl.* **58**, 1207–1216 (2016).

73. Wang, Y. K. *et al.* Rigidity of Collagen Fibrils Controls Collagen Gel-induced Down-regulation of Focal Adhesion Complex Proteins Mediated by α2β1 Integrin. *J. Biol. Chem.* **278**, 21886–21892 (2003).
**Figure. 1**. Schematic overview of the research.
**Figure. 2**. The schematic of the microfluidic system. Its dimensions are introduced as parameters, which are explained in the simulation section. This device is composed of two lateral media channels and one interposed gel channel.
**Figure. 3.** Schematic of microfluidic chip fabrication using photolithography, collagen solution loading process without any leakage, and cells encapsulated in the collagen3-BGNs3 hydrogel into the chip.

**Figure. 4.** Simulation of gel filling into the microfluidic device for two different designs (a) Movement of collagen solution without leakage along the gel channel of the design number 3; (b) Movement of collagen solution without leakage along the gel channel of the design number 2.

**Figure. 5.** Simulation of gel filling into the microfluidic device for different viscosities: (a) Viscosity of 9 mPa×s and low leakage of collagen solution along the gel channel; (b) Viscosity of 11 mPa×s and leakage of collagen solution and entering the media channel.

**Figure. 6.** Mass transfer simulation within the gel channel of the designed chip (Design Number 2). (a) t = 2.5 h (b) t = 4 h.

**Figure. 7.** Characterizing collagen type I, derived from rat tail. (a) Freeze-dried collagen; (b) the collagen solution, pre-crosslinking at a concentration of 3 mg/mL; (c) self-assembled collagen under pH 7.4 and 37 °C conditions; (d) The FTIR spectrum indicating specific molecular vibrations such as amide A, amide B, amide I, amide II, and amide III, confirming the presence of the type 1 collagen structure; (e) SDS-PAGE data revealed collagen type 1 with two $α_1$ chains and one $α_2$ chain, as well as a β-dimer, further confirming the structure of collagen type I.

**Figure. 8.** The characterization results of the synthesized BGNs with the sol-gel method (a) XRD spectra; (b) General observation photographs; (c) FTIR spectra; (d) FESEM micrograph image of BGNs; (e) Particle size distributions.; (f) EDX analysis of BGNs.

**Figure. 9.** SEM images of collagen hydrogels with varying concentrations of BGNs components. (a). collagen3-BGNs0; (b, c). Collagen3-BGNs1; (d). Collagen3-BGNs2; (e, f). Collagen3-BGNs3; Yellow arrows indicate BGNs. **(g)** Rheological evaluation of storage and loss modulus of different hydrogels as a function of strain at the same temperature.

**Figure. 10.** Evaluation of cell viability encapsulated in hydrogels. (a) Comparison of the cell viability in different concentrations of BGNs by MTT assay; (NS, no significant difference; $p ≤ 0.05$, *** $p ≤ 0.001$, **** $p ≤ 0.0001$); (b) Live/dead was utilized to assess the viability of the cells encapsulated in collagen3-BGNs3 hydrogel into the microfluidic on day 3, the cells were highly viable, the green color shows the living cells, and the red color indicates the dead cells (scale bar = 200 μm).

**Table.1.** The dimensions of the trapezoids used in the simulation.

**Table.2.** Governing equations' parameters.

**Table.3.** Description of the hydrogel composite samples.

## Author Contributions

**Faezeh Ghobadi**: Conceptualization, Methodology, Formal Analysis, Writing Original Draft. **Maryam Saadatmand**: Supervision, Project Administration, Reviewing and Editing. Resources. **Sara Simorgh**: Resources, Reviewing and Editing. **Peiman Brouki Milan:** Resources.

## Additional Information

### Competing Interests
The authors declare no competing interests.